\documentclass[%
 reprint,12pt
 amsmath,amssymb,
 onecolumn,notitlepage
]{article}
\usepackage{mathtools}
\usepackage{amsfonts}
\usepackage{bm}
\usepackage{authblk}
\usepackage{graphicx}
\usepackage{dcolumn}%
\usepackage{bm}
\usepackage[euler]{textgreek} 
\usepackage{color} 
\usepackage{subcaption}
\usepackage{esvect}
\usepackage{cite} 
\usepackage{hyperref}
\usepackage[a4paper, total={5.5in, 7.5in}]{geometry}
\hypersetup{
    colorlinks,
    citecolor=blue,
    filecolor=black,
    linkcolor=blue,
    urlcolor=blue
}
\usepackage[dvipsnames]{xcolor}
\usepackage{titlesec}
\titleformat*{\section}{\large\bfseries}
\titleformat*{\subsection}{\bfseries}
\usepackage{amsmath,amssymb}
\begin{document}

\title{Compact Itinerant Microwave Photonics with \\ Superconducting High-Kinetic Inductance Microstrips}

\author[1]{Samuel Goldstein}
\author[1]{Guy Pardo}
\author[1]{Naftali Kirsh}
\author[2]{Niklas Gaiser}
\author[2]{Ciprian Padurariu}
\author[2,3]{Bj{\"o}rn Kubala}
\author[2]{Joachim Ankerhold}
\author[1]{Nadav Katz}

\affil[1]{The Racah Institute of Physics, The Hebrew University of Jerusalem, Givat Ram, Jerusalem 91904, Israel}

\affil[2]{Institute for Complex Quantum Systems and IQST, University of Ulm, 89069 Ulm, Germany}

\affil[3]{Institute of Quantum Technologies, German Aerospace Center (DLR), 89077 Ulm, Germany}

\date{\today}
\renewcommand\Affilfont{\itshape\small}
\maketitle
\begin{abstract}
Microwave photonics is a remarkably powerful system for quantum simulation and technologies, but its integration in superconducting circuits, superior in many aspects, is constrained by the long wavelengths and impedance mismatches in this platform. We introduce a solution to these difficulties via compact networks of high-kinetic inductance microstrip waveguides and coupling wires with strongly reduced phase velocities. We demonstrate broadband capabilities for superconducting microwave photonics in terms of routing, emulation and generalized linear and nonlinear networks.
\end{abstract}

\section{Introduction and motivation}
Itinerant optical photonics \cite{dmitriev2017quantum,wang2016experimental,zhong2020quantum,yang2021quantum} was made possible by the low loss, short wavelength, and controlled patterning in optical on-chip devices, enabling multimode interferometry. Aside from demonstrating superposition and multi-partite entanglement, these systems are proposed as a path to quantum-technological applications \cite{o2009photonic,pirandola2018advances,wang2020integrated}. A clear and persistent disadvantage of these devices is the challenge of on-demand single optical photon generation\cite{chang2014quantum}. 
 
In contrast, superconducting circuits demonstrated high quality, on-demand single microwave photons more than a decade ago  \cite{houck2007generating}. As superconducting qubit systems emerge as a leading candidate in the race towards universal quantum computing, it is vital to integrate microwave photonics for routing, processing and communication between computational nodes \cite{huang2020superconducting,flurin2012generating,juliusson2016manipulating}.  

The ubiquitous frequencies of microwave quantum circuits are constrained between $\sim~10^9$-$10^{10}$~Hz \cite{gao2008physics,malyshev2020microwave} due to a combination of fundamental and technical considerations \cite{krantz2019quantum}. This leads to typical wavelengths, $\lambda$, in excess of 10~mm and enlarged overall device sizes, with consequent box-mode parasitic excitations and fabrication difficulties when trying to scale to complex networks of microwave photonics \cite{lienhard2019microwave,pozar2011microwave}. It has been suggested to compress footprints by deforming the traces to spirals or meanders \cite{morvan2021bulk,bockstiegel2014development}. However, such elongated devices are more vulnerable to fabrication errors leading to "weak spots" \cite{adamyan2016superconducting} and increased noise from magnetic vortex penetration \cite{koch2007model}.

The high-kinetic inductance (HKI) of amorphous superconductors (such as WSi) along with a large microstrip capacitance introduced in this work, allows us to achieve impedance-matched short wavelength microwave photonics. This fulfills the linear networking properties considered above. In addition, the nonlinearity \cite{kher2016kinetic,valenti2019interplay} of such HKI microstrips provides a route to amplification at the quantum limit \cite{eom2012wideband,bockstiegel2014development,adamyan2016superconducting,vissers2016low,chaudhuri2017broadband} due to wave-mixing  phenomena. Single HKI superconducting strips have been used to build high-quality microwave resonators \cite{basset2019high}, superinductors for use in qubit architectures\cite{gruenhaupt2019design}, kinetic inductance detectors \cite{valenti2019interplay,day2003broadband,mazin2010thin,kerman2006kinetic}, galvanometers \cite{doerner2018compact}, and more. This can now be extended to a multi-mode network for more complex photonic tasks.

Here we establish a scalable platform for itinerant microwave networks by demonstrating a variety of geometries of superconducting HKI WSi coupled microstrips. We achieve controllable links between the $50~\Omega$ impedance-matched central traces by using sub-micronic \cite{niepce2019high} coupling traces.

This paper is structured as follows: In Section \ref{sec:theory} we briefly review the concept of high-kinetic inductance, show how we adopt the microstrip geometry for coupled superconducting networks, 
and derive the theory of propagation along microstrips connected periodically by coupling traces. For our simplest device this leads to a two-mode bandstructure calculation. This section also describes the simulation of our devices for an arbitrary geometry. Section \ref{sec:Fabrication} describes the fabrication of our devices, including design choices made in anticipation of the physical phenomena we want to observe, and the technical recipe for fabrication (extended details are provided in Appendix~\ref{Sec:TechnicalFab}). In Section \ref{sec:ResultsDiscussion} we present the results of experiments with networks of traveling and standing waveguides, and we discuss their linear and nonlinear behavior. Finally, Section~\ref{sec:Outlook} considers various applications of our findings. Appendices contain more information on the phase velocity's experimental value, the technical details of the fabrication, the Fabry-Perot resonances in the coupled mode formalism, and specific supporting simulations.

\begin{figure*}[t!]
\includegraphics[clip,trim=1cm 7.7cm 0cm 2cm,width=1\textwidth]{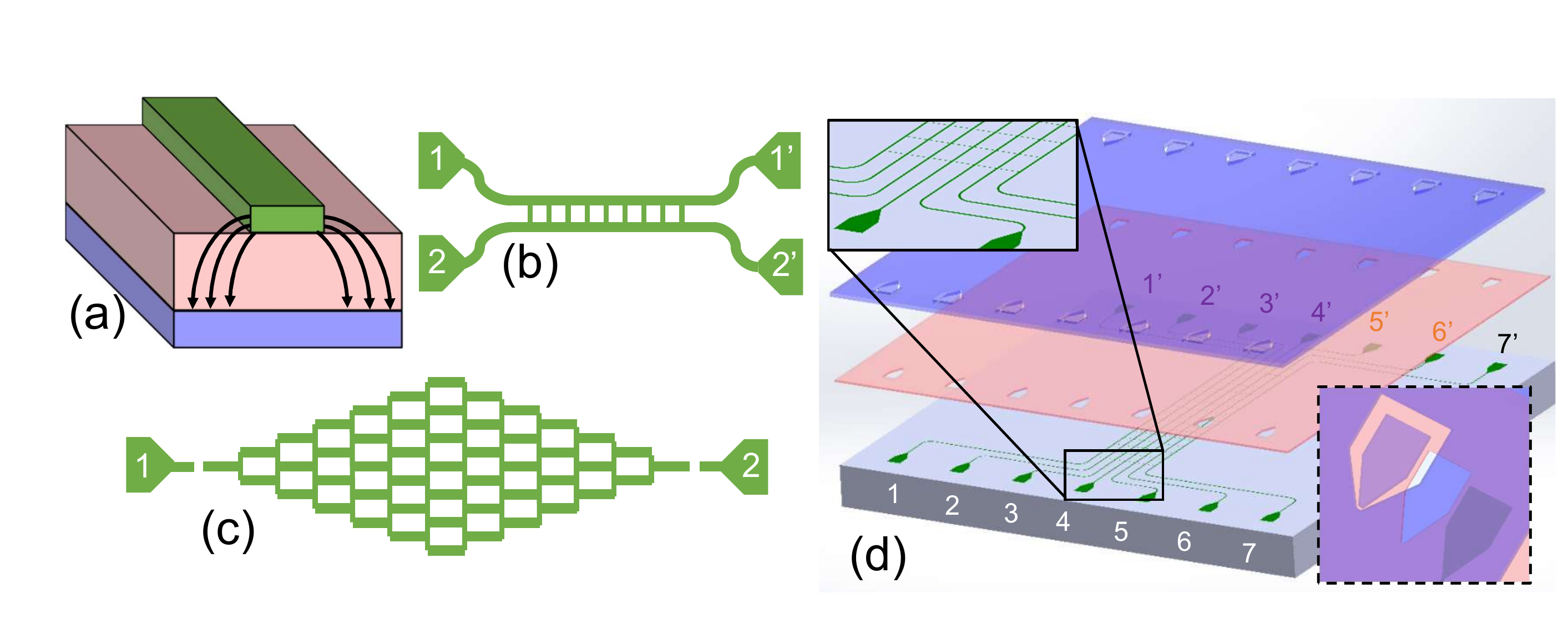}
\caption{Graphic representation of the microstrip devices. (a) Conceptual microstrip trace geometry; arrows indicate the electric field. Green: superconducting WSi trace, pink: dielectric, purple: Al ground. (b) The double-line device: Two periodically coupled traveling waveguides. (c) The resonant lattice structure, 49 nearest-neighbor coupled standing waveguides, with additional capacitive couplings to launchers in either end. (d) Illustration of the generalization of (b) to seven parallel microstrip traces (the 7PMT), also showing the various layers of fabrication. Note the inversion compared to (a); in our fabrication scheme the WSi is deposited first, and the ground last. Top-left corner inset: Zoom, showing traces and coupling lines. Bottom-right inset: Angled top view, showing the opening in Al and Si layers, fitting the launch pad of the WSi seen as shadow.}
\label{fig:Concept}
\end{figure*}

\section{Theoretical Framework}
\label{sec:theory}

\subsection{High-Kinetic Inductance Waveguides}
The kinetic inductance in superconducting devices operated in the microwave regime stems from the the kinetic energy per unit length associated with the motional energy of the Cooper pairs in the device \cite{meservey1969measurements}. The kinetic energy of a single Cooper pair is $\frac{1}{2}(2m_e)v_s^2$, where $m_e$ is the electron mass and $v_s$ is the velocity. The density of pairs equals half the density of electrons $n_e$, so the kinetic energy per unit length can be written 
\begin{equation}
    E_k=\frac{1}{2}(2m_e) v_s^2\cdot \frac{n_e}{2} A=\frac{m_e}{2q_e^2 n_e A}I^2
\end{equation}
where $A$ is the cross section. Also $I=q_e n_e v_s A$ is the current, where $q_e$ is the elementary charge. $E_k$ is thus added to the energy of the magnetic field $E_m$ induced when the charge carriers are set in motion \cite{mallory1997electrically}. We use the common definition of the total inductance per unit length as related to the total energy due to current (also per unit length) by $E_k+E_m=\frac{1}{2}L_l I^2$. Thus the $L_l$ of a superconducting transmission line is comprised of the sum of the kinetic term and the magnetic contribution \cite{sypkens2021development}:

\begin{equation}
    L_l=\frac{\mu_0\lambda_L^2}{A}\bigg(1+\Big(\frac{I}{I_\star}\Big)^2+\cdots\bigg)+L_{g,l}
    \label{eq:Lkin}
\end{equation}
where $\mu_0$ is the vacuum permeability, $\lambda_L$ the London penetration depth ($\sim 450$ nm for our WSi traces), and is given by $\lambda_L^2 =m_e/\mu_0 n_e q_e^2$ \cite{kittel1996introduction}. In this expression, we have also added $I_\star$ as the characteristic current scale for nonlinearity \cite{eom2012wideband, malnou2021three}. The geometric (magnetic) inductance per unit length, $L_{g,l}$ is typically negligible compared to the HKI (first term in $L_l$) for our thin WSi traces \cite{marychev2021extraordinary}. 
 
 The nonlinear kinetic inductance (given by the factor $(I/I_{\star})^2$ in Eq.~\eqref{eq:Lkin}) becomes relevant as larger currents are driven through the traces ($\lesssim I_\star$). The underlying physics explaining this nonlinearity stems from a perturbative suppression of the superconducting order parameter as the current is increased \cite{semenov2021superconducting}, and consequently the pair density is suppressed, resulting in a quadratic (nonlinear) increase in $\lambda_L$ affecting the prefactor of Eq.~\eqref{eq:Lkin}.  In our devices, typically $I_\star \simeq 3 I_c$, where $I_c$ is the critical current of the traces \cite{semenov2020effect}.
 
We note that the kinetic inductance is a function of transport properties of the superconducting material (especially the large penetration depth $\lambda_L$) as well as geometric parameters (such as the small cross-section $A$). 

\subsection{Superconducting microstrips}

We achieve characteristic impedance-matching in our microstrips with relative ease. The capacitance per unit length $C_l=\varepsilon_r \varepsilon_0 w/d$ can be engineered to fit $L_l$ to reach the impedance $Z = \sqrt{L_l/C_l} = 50$. Here, $\varepsilon_r$ and $d$ are the dielectric constant and the thickness of the dielectric layer, and $\varepsilon_0$ the vacuum permittivity. This contrasts the case of coplanar HKI traces, where extended tapers are required to avoid reflections due to discontinuity in $Z$ \cite{klopfenstein1956transmission}. The use of impedance-matched microstrips makes the tapers superfluous, reducing the area further. Microstrip traces are essentially parallel plate capacitors, with transverse electromagnetic fields penetrating the dielectric material separating the ground plane from a conducting trace as displayed in Figure 1(a). When connected directly to larger wire-bond launchers (e.g. our ”double-line” device, depicted in Figure \ref{fig:Concept}(b)), the microstrips constitute traveling waveguides. Alternatively, standing waveguides, can be implemented with microstrips, when the traces are open or grounded at either end.

Transmission between adjacent microstrips is achieved through sub-micronic coupling wires (”couplers”). As the couplers’ widths are narrowed down to about 1/10 of the 50~$\Omega$ waveguides’ width, $L_l$ of the former is increased by an order of magnitude. Following these geometric changes also $C_l$ changes its value to become smaller by the same ratio. Thus the couplers behave as mostly inductive links. The couplers’ $Z_l$ is therefore an order of magnitude larger than that of the waveguides, confirming their perturbative role as a weak link.

A fundamental advantage of the microstrip architecture is the very slow phase velocity (approximately 1\% of the vaccuum speed of light), 
\begin{equation}
v_{ph}=c \Bigg(\varepsilon_r\bigg(1+\frac{\lambda_1}{d}\coth\frac{t_1}{\lambda_1}+\frac{\lambda_2}{d}\coth\frac{t_2}{\lambda_2}\bigg) \Bigg) \simeq \frac{1}{\sqrt{L_lC_l}}
\label{eq:v_ph}
\end{equation}
 where $\lambda_{1,2}$ and $t_{1,2}$ are the superconducting penetration depths and thicknesses of the two superconductors; the trace and the ground \cite{mazin2010thin}. In the devices presented in this paper $v_{ph}\simeq 4\times10^6$. The immediate consequence for traveling waves (cf. in the amplifier in \cite{goldstein2020four} and in the first two devices shown here) is that the photons are decelerated to spend several nanoseconds in our device, permitting us to shorten the traces significantly and still maintain sufficient wave-mixing or appreciable routing to other coupled waveguides. In the case of resonant structures, waveguide lengths' $L$ can be reduced according to
$L=\lambda/2 \sim v_{ph} / 2f$ where $f$ is the desired frequency, cf. the operational bandwidth. For our $v_{ph}$ and $f$'s this corresponds to $L\sim 200$ \textmugreek m.

\subsection{Theory of periodically coupled traveling waveguides}

Aiming towards functionalization, HKI microstrip networks, which form periodic one- or two-dimensional structures, are of obvious interest. Here, we present the (linear) theory of two periodically coupled waveguides of infinite length described and analyzed in the language of crystal physics. This approach is easily extendable to multi-trace networks or two-dimensional devices and can serve as starting point for more advanced descriptions including nonlinearities and quantum effects.

\begin{figure*}[t!]
\includegraphics[clip,trim=6cm 2cm 6cm 2cm,width=1\textwidth]{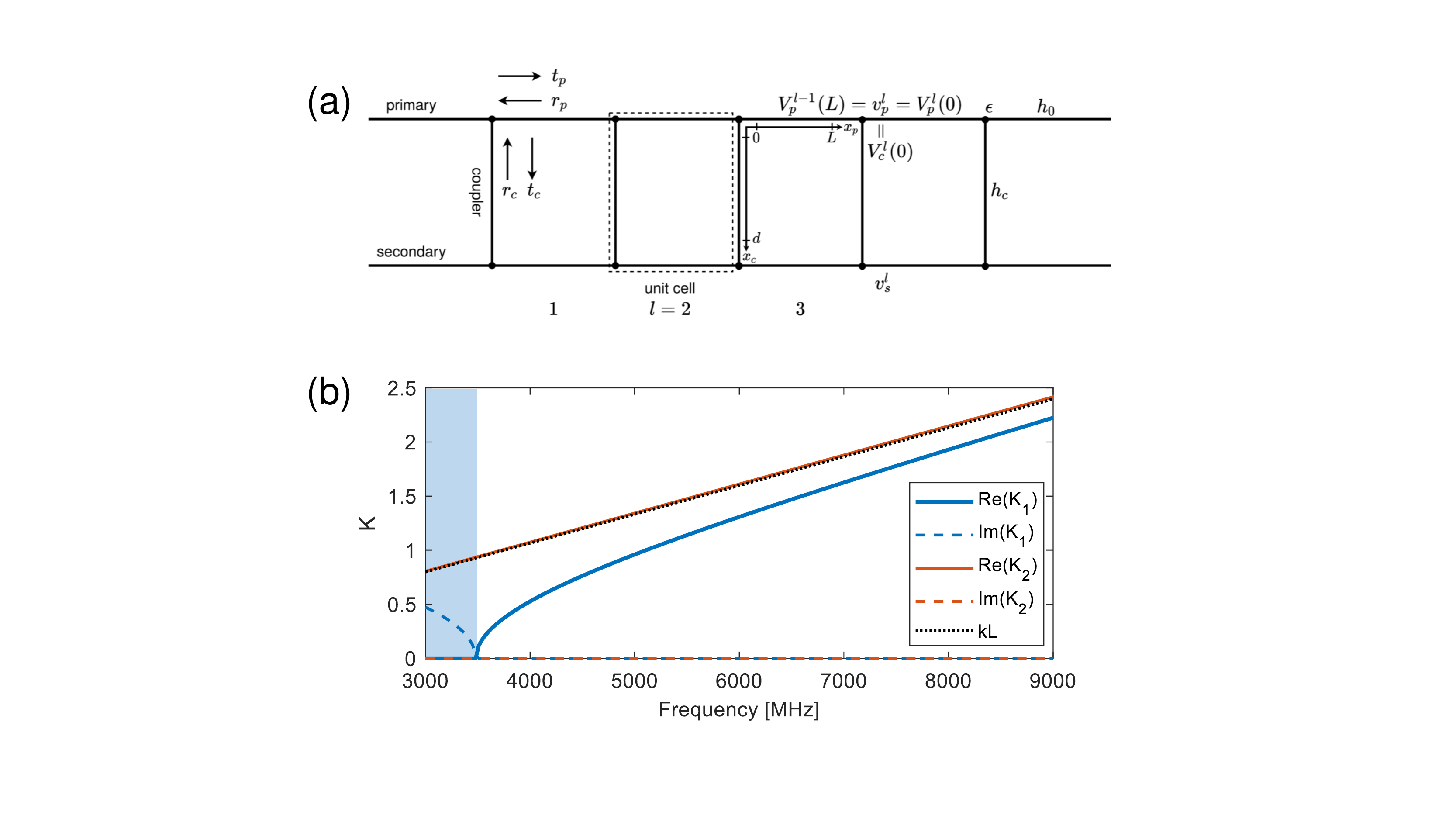}
\caption{\label{fig:double-line_defs} (a)  Sketch of the double-trace structure and definitions introduced in the text. Using the voltages on the nodes $v^l_\alpha$ as variables, Kirchhoff rules yield a tight-binding model with parameters $\epsilon,h_{0,c}$ as marked on the rightmost unit cell. (b) Bandstructure of a periodic infinite double-line device (parameters adapted from the experimental device). 
The (Bloch-)wavevector $K_2\approx kL$ of the eigenmode symmetric in primary and secondary line is nearly unaffected by the couplers, while the anti-symmetric mode has a band gap (shaded region) where $\textrm{Im}\,K_1>0$.
\\
}
\end{figure*}

A standard transmission line model (see e.g, Reference~\cite{pozar2011microwave}) yields wave propagation along the various segments of the structure,
\begin{align}
    V_\alpha^l(x_\alpha)   &=  t_\alpha^l e^{ i k_\alpha x_\alpha} +     r_\alpha^l e^{- i k_\alpha x_\alpha} \quad \text{where}\;  \alpha= p,s,c\,;\quad k_\alpha = 2\pi f \sqrt{L_\alpha C_\alpha} \\
   Z_\alpha  I_\alpha^l(x_\alpha)   &=   t_\alpha^l e^{ i k_\alpha x_\alpha} -    r_\alpha^l e^{ -i k_\alpha x_\alpha}   \quad  \text{and}\;   Z_\alpha = \sqrt{L_\alpha / C_\alpha}\,. 
\end{align}
The different segments are distinguished by an index $l$ numbering the unit cells and $\alpha=p,s,c$ for primary, secondary and coupler lines, see Figure~\ref{fig:double-line_defs}(a), where for the designed double-line, we can assume $Z_p = Z_s =: Z_0,\;k_p=k_s=:k_0$, while $x_{p/s} \in [0,\,L]$ and $x_c\in[0,\,d]$. Dissipation can also easily be included.
Kirchhoff circuit equations require voltage matching and current conservation for each node, e. g.,
\begin{subequations}\label{voltage nodes}
 \begin{align}
 t_p^{l-1} e^{ i k_p L}  + r_p^{l-1} e^{- i k_p L} &= V_p^{l-1}(L) \equiv v_p^l =   V_p^{l}(0) = V_c^{l}(0) \\
0 &= I_p^{l-1}(L)   - I_p^{l}(0) - I_c^{l}(0)  \,, 
\end{align}
\end{subequations}
where the first line can be used to rewrite the current in each segment in terms of two voltage node variables it connects. This immediately yields a tight-binding description
\begin{subequations}\label{eq:tight-binding}
 \begin{align}
 0  =  h_0 v_p^{l-1} + \epsilon v_p^{l} + h_0  v_p^{l+1} + h_c  v_s^{l} \\
 0  =  h_0 v_s^{l-1} + \epsilon v_s^{l} + h_0  v_s^{l+1} + h_c  v_p^{l}
\end{align}
\end{subequations}
 with real parameters for on-site energy and in- and cross-line hoppings,
\begin{equation}
\epsilon = i \left( \frac{2}{Z_0} \frac{1 + z_0^2}{1- z_0^2} + \frac{1}{Z_c} \frac{1 + z_c^2}{1- z_c^2} \right) \quad \text{and}\quad 
h_{0/c} = -i \frac{1}{Z_{0/c}} \frac{2 z_{0/c}}{1- z_{0/c}^2} \;,
\end{equation}
where  $z_0=e^{ i k_0 L}$ and  $z_c=e^{ i k_c d}$ were introduced.

The eigenmodes of an infinite double-line are straightforwardly found by a Bloch-like ansatz
\begin{align}
\left(
\begin{array}{c}
  v_p^l  \\
  v_s^l   
\end{array}
\right) =
\vec{v}_\nu 
e^{i K_\nu l} \quad;\;\; K_\nu \in \mathbb{C}\;, \nu=1,2
\end{align}
as eigensolutions of 
\begin{align}
\left(
\begin{array}{cc}
 \epsilon + 2 h_0 \cos K_\nu  & h_c  \\
  h_c & \epsilon + 2 h_0 \cos K_\nu
\end{array}
\right) \;
\vec{v}_\nu   =0 \quad \Rightarrow \cos K_{1/2} = - \frac{\epsilon}{2 h_0} \pm \sqrt{\frac{h_c^2}{4 h_0^2} }, \end{align}
see Figure~\ref{fig:double-line_defs}(b), with eigenvectors that are (anti-)symmetric in primary and secondary voltages for two identical coupled lines.

To solve the actual scattering problem for a finite line we write the ansatz for the general solution
\begin{equation}
\begin{pmatrix}
v_p^l  \\
  v_s^l  
\end{pmatrix}
= \sum_{\nu=1,2} a_\nu \vec{v}_\nu e^{i K_\nu l} +  b_\nu \vec{v}_\nu e^{-i K_\nu l}  
\end{equation}
and use Equation~\eqref{voltage nodes} to express the left/right-going amplitudes of a unit cell by node voltages
\begin{subequations}\label{voltage nodes b}
 \begin{align}
 t_\alpha^{l-1} &= \frac{1}{1-z_0^2} \left(  v_\alpha^{l-1}   - z_0 v_\alpha^l \right)  \\
 r_\alpha^{l-1} &= \frac{1}{1-z_0^2} \left(  -z_0^2 v_\alpha^{l-1}   + z_0 v_\alpha^l \right) \qquad ,\text{where} \quad \alpha=p/s \,,
\end{align}
\end{subequations}
where the input and output of a line with $N$ nodes is
\begin{equation} \label{in_out_amplitudes}
t^L_\alpha = t^0_\alpha\,,\quad r^L_\alpha = r^0_\alpha\,,\quad
t^R_\alpha = z_0 t^N_\alpha \,,\quad r^R_\alpha = \bar{z}_0 r^N_\alpha\,.
\end{equation}
The four variables in the ansatz are then determined by four of the eight equations, Equation.~\eqref{voltage nodes b},  involving the known boundary conditions (e.g., the inputs into all lines), while the remaining four equations yield the unknown output variables.

In that manner, one may, for instance, find for the case of a single input, $t^L_p$ into the primary line (and all other inputs set to zero),
\begin{equation}\label{t_p/s}
t^R_{p/s} = \frac{t^\textrm{tot}(K_2) \pm t^\textrm{tot}(K_1)}{2} t^L_p\;.
\end{equation}
with 
\begin{equation}\label{t_tot_result}
t^\textrm{tot}(K_\nu) = \frac{(1-z_0^2) \sin{K_\nu}}{(1/z_0)\sin{[K_\nu(N+1)]} -2 \sin{[K_\nu N]} +z_0 \sin{[K_\nu(N-1)]}} \,.
\end{equation}
We will explain the specific form of Eqs.\eqref{t_p/s},\eqref{t_tot_result} in the discussion of the results below.

\subsection{Simulation}

Besides the band theory for periodic devices explained above, which can give analytical results for the simplest cases, we apply a number of numerical simulations to model different aspects of the physics of the various investigated devices on varying levels of complexity. Here, we describe a generic method usable for arbitrary linear networks and comment how some nonlinear effects can be accounted for, while other more specific approaches are briefly explained in the appendices.

Defining the network geometry, we consider $q=1,\hdots,Q$ nodes, some of which are connected by edges. Besides its length $\delta_j$, each edge $j=1,\hdots, J_q$ connected to node $q$ is characterized by capacitance and inductance per unit length (determined, e.g., by different widths of main traces and nanowire couplers) yielding an impedance $Z_j$.
The edge voltage  then can be written (in line with Equation~\eqref{voltage nodes}) as
\begin{equation}
    V_q(x)=A_q^j e^{ik_j x}+B_q^je^{-ik_jx},
    \label{eq:voltage}
\end{equation}
where $x$ denotes the distance from the $q$'th node along the $j$'th edge and $k_j$ is the impedance and frequency dependent wave-number. We solve for $A_q^j$ and $B_q^j$ for all $Q$ nodes' $J=\sum_{q}J_q$ connected edges, but the number of unknowns can be reduced by mapping the connectivity: 
\begin{equation}
    A_{q_1}^j=B_{q_2}^j e^{-ik_j\delta_{j}}
    \label{eq:contin}
\end{equation} if the $j$'th edge connects the nodes indexed $q_1$ and $q_2$. 
The input and output nodes (injection and readout) constitute the boundary conditions. For all other nodes current conservation requires that
\begin{equation}
     \sum_{j=1}^{J_q} \frac{A_q^j}{Z_j} =\sum_{j=1}^{J_q} \frac{B_q^j}{Z_j}.
     \label{eq:current}
\end{equation}
We encode Eqs. \eqref{eq:voltage}-\eqref{eq:current} together with the boundary conditions in a matrix, \textbf{M}, in which each row represents an equation, so that 
\begin{equation}\label{eq:matrixeq}
    \textbf{M} \times \vv{V} = \vv{K}
\end{equation}
where $\vv{V}=(A_1^1,B_1^1,A_1^2...A_1^{J_q},B_1^{J_q}...A_2^1,B_2^1...A_Q^1,B_Q^1...A_Q^{J_Q},B_Q^{J_Q})$, i.e. the vector of unknowns, and $\vv{K}$ is a vector almost exclusively of zeros due to the nature of the equations, except for those regarding the boundary conditions. Our simulation also accounts for dielectric losses as we add an imaginary term to $k_j$.

Accounting for the nonlinear inductance of our devices leads to a power-dependent wave-equation for each edge, which, in general, yields complicated frequency mixing physics (as exploited for traveling-wave parametric amplifiers \cite{goldstein2020four,erickson2017theory}). 
Here, we will not consider those effects, but solve the nonlinear partial differential equation within a single-frequency ansatz (namely with the frequency of the CW-input). Thereby, it reduces to coupled ordinary differential equations (ODEs)  for $V(x)$ and $I(x)$ for each segment of our device. In that nonlinear case, we can still encode voltage and current at the nodes (i.e. at the end points of each segment) by amplitudes $A_q^J$ and $B_q^j$, but the propagation along the segment and the relation between amplitudes at start and end is no longer trivially given by the phase factor of a propagating wave, Equation~\eqref{eq:contin}, but rather has to be found by solving the corresponding ODEs for each segment. This means that if nonlinear effects are included for a single segment $j$ between nodes $q_1$ and $q_2$, the two lines in the matrix equation Equation~\eqref{eq:matrixeq} corresponding to Equation~\eqref{eq:contin} (and the corresponding equation linking $B^J_{q1}$ to $A^J_{q2}$) are replaced by a nonlinear relation between the four amplitudes, which is implicitly defined by solving the corresponding ODEs. The matrix Equation~\eqref{eq:matrixeq} thus becomes a nonlinear implicit equation.

\section{Fabrication}
\label{sec:Fabrication}

In designing our devices, we consider different aspects directly controlled by the dimensions of the traces. Once the width $w$ and height $t$ of the microstrip are chosen, $L_l$ is settled given its material properties, and the requirement of impedance matching determines the dielectric layer thickness necessary to reach the proper value of $C_l$.

\begin{figure*}[t!]
\includegraphics[clip,trim=0cm 10cm 0cm 6cm,width=1\textwidth]{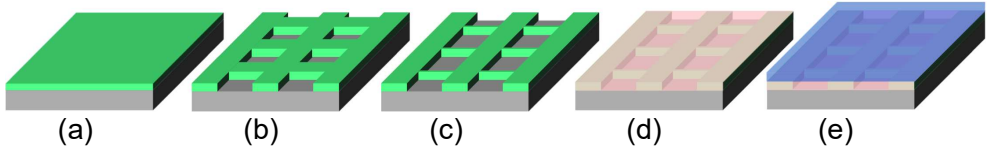}
\caption{Graphic illustration of the fabrication steps (not in scale). (a) Sputtering of WSi (green) covers the entire surface of the bulk Si substrate (gray). (b) Optical lithography and wet-etch define the network traces. (c) Electronic lithography narrows down the couplers (horizontally oriented in this chart) to the desired submicronic width. (d) aSi (pink) is deposited onto the WSi network, and (e) the Al (ground) is finally added, also by e-beam deposition. The latter layer is subsequently patterned by liftoff (not shown).}
\label{fig:FabFlow}
\end{figure*}

A central concern of the design is to ensure step coverage. The dielectric layer must necessarily be thicker than the underlying patterned WSi network; when the opposite is the case, the dielectric layer fails to climb and cover the edges of the network, allowing electrical shorts to the ground layer. This constraint disqualifies the use of certain dielectrics, e.g. SiO$_2$, with relatively low $\varepsilon_r$.

All our devices are fabricated by five consecutive steps to define their three layers.
Initially, we grow a $\sim 10$~nm film of WSi by DC magnetron sputtering, where the stoichiometry of the target (45\%/55\%) together with the dimensions of the trace, yet to be defined, determines $\lambda_L$ and hence $L_{kin}$. A protective resist mask is then applied, first by spinning, and next by optical lithography allowing wet-etch of WSi everywhere except on the intended network segments. Electronic lithography is used to narrow the coupler-width from the scale of $\sim $\textmugreek m, where optical lithography is efficient, to $\sim 100$~nm, below the wavelength of our laser-writer, again by wet-etch. Next, the dielectric is grown at a rate of $\sim$0.1 nm$/$s by e-beam evaporation without further patterning. The inclusion of dielectrics potentially results in loss effects, considered further below. We choose amorphous silicon with the dielectric constant $\varepsilon_r \simeq 11.7$ for this purpose \cite{pozar2011microwave}.

In the last fabrication step, we prepare a double-layer photo-resist mask. The lower layer's enhanced sensitivity to the laser compared to the upper layer, results in an "undercut", ensuring a smooth liftoff in acetone after evaporation of the Al top film. The Al serves as the electrical ground of the microstrips and protects the device mechanically during continued handling. After dicing into $6\times6$ mm$^2$ squares, wire-bonding to impedance-matched printed circuit boards, and mounting in Al boxes, all experiments are conducted at $\simeq 20$~mK temperatures in our dilution refrigerator, far below WSi's critical temperature of $4.7$ K \cite{seleznev2016superconducting}.

\section{Results and Discussion}
\label{sec:ResultsDiscussion}
\begin{figure*}[t!]
\includegraphics[clip,trim=4cm 5cm 8cm 6.5cm,width=1\textwidth]{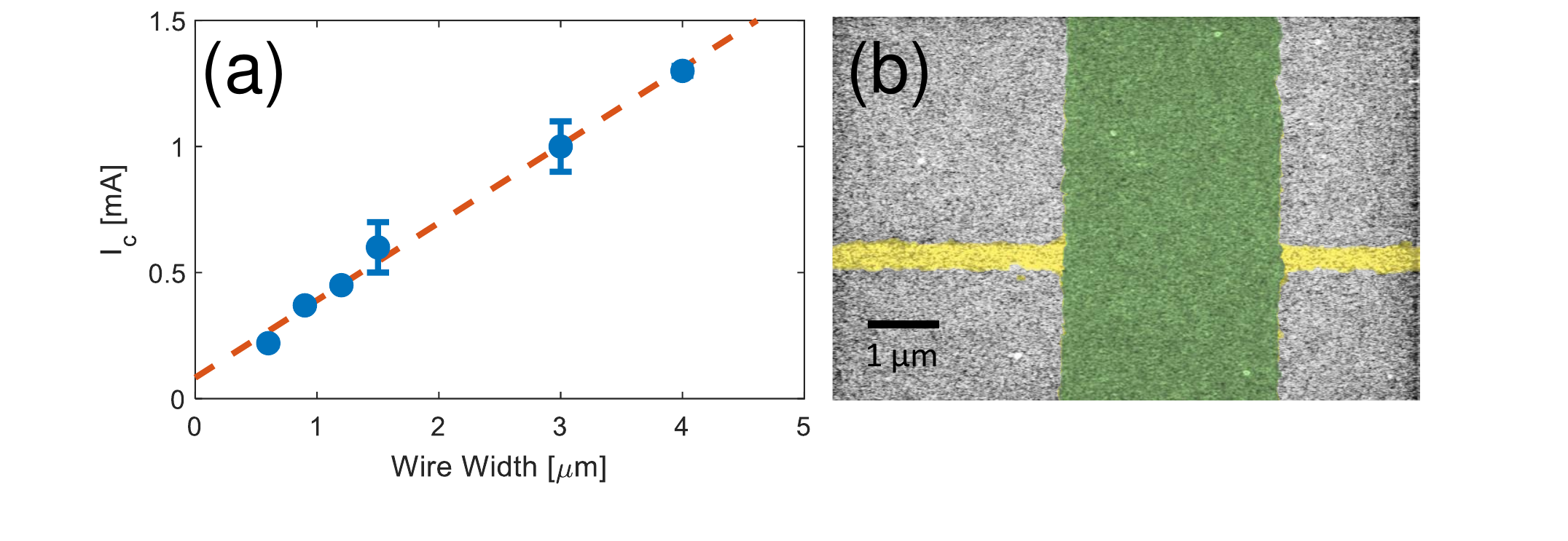}
\caption{(a) Critical current measurement of a $\sim$12 nm thick WSi chip with nanowires of various widths. The dashed line is a linear fit. Reprinted with permission from Reference \cite{kirsh2021linear}. \copyright~2021 by the American Physical Society. (b) SEM photo (false-colored)  from the 7PMT device of a 3 \textmugreek m wide waveguide (green) and an intersecting submicronic coupler (yellow) connecting the waveguide to parallel waveguides.}
\label{fig:WSiWireCharacterize}
\end{figure*}

Before proceeding to observing the behavior of couplers in networks, we measure their critical currents and find $I_c\simeq 0.15$ mA, which is consistent with the scaling of critical currents with width, found for wider superconducting WSi traces shown in Figure~\ref{fig:WSiWireCharacterize}(a) \cite{kirsh2021linear}. This linear scaling of the critical current  with trace cross section emphasizes the advantage of our fabrication method; the e-beam lithography ensures accurate dimensions of couplers and waveguides (exemplified with a SEM photo in Figure~\ref{fig:WSiWireCharacterize}(b)) and yields the desired nonlinearity.

\subsection{Networks of Traveling Waveguides}
\label{subsec:TravellingResults}
Our first device, is a "double-line", i.e. two parallel $3~$\textmugreek m wide microstrips, separated by $30$~\textmugreek m, and connected every $100$~\textmugreek m by 30~couplers. This periodicity ensures mode coupling under the slowly varying envelope approximation considering the reduced $v_{ph}$.

We measure the output from both lines (ports 1' and port 2'), when continuous waves (CW) signals are applied from our Keysight P5024A Vector Network Analyzer into one of them (see Figure~\ref{fig:Concept}(b)). The total length of each microstrip, i.e. from launcher to launcher, is $3$ mm $> \lambda \simeq 400$~\textmugreek m. The unemployed launcher (port 2 in Figure~\ref{fig:Concept}(b)) is terminated to the ground through attenuators and a $50\Omega$ resistor at room temperature to avoid reflections into the waveguide. 

\begin{figure*}[t!]
\includegraphics[clip,trim=1.5cm 8cm 1cm 4cm,width=1\textwidth]{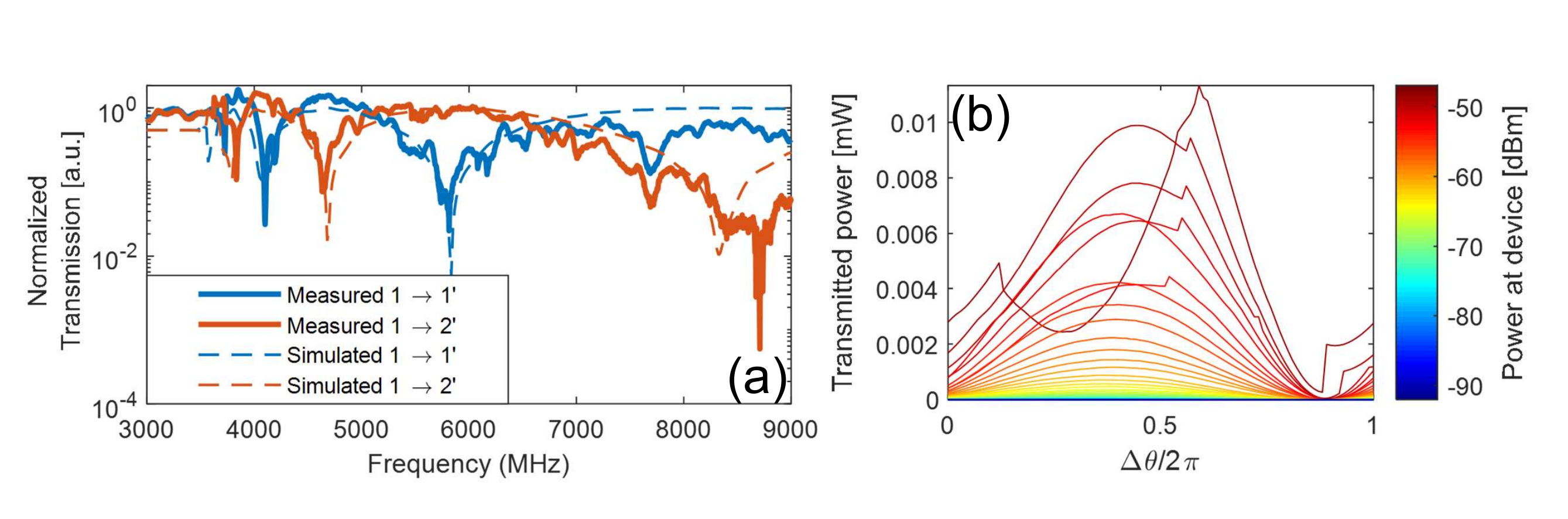}
\caption{(a) Direct and coupled transmission measurements and simulation for two parallel microstrip traces periodically coupled through highly inductive nano-wires. (b) Phase-dependent transmission, as CW signals are applied in both traces simultaneously, with changing input powers in port 1 (represented by colorbar) and changing phase in port 2 (horizontal axis). The measured output in port 2' (vertical axis).}
\label{fig:Nockit2}
\end{figure*}

The observed frequency-dependent transmission, Figure~\ref{fig:Nockit2}(a), shows a flat region at low frequencies, where both, direct and coupled, lines transmit well, followed by a series of resonances (anti-resonances) in the direct transmission with concomitant anti-resonances (resonances) in the coupled transmission.

This behavior is well reproduced by simulations of the circuit (dashed) based on voltage continuity and current conservation. 
To analyze the results, we first consider the eigenmodes of an infinitely extended tight-binding model, see Equation~\eqref{eq:tight-binding}. The band structure of the two resulting symmetric and anti-symmetric eigenmodes is shown in Figure~\ref{fig:double-line_defs}(b), showing a symmetric mode which propagates with a Bloch-wave vector $K\approx kL$ in the probed frequency range while the antisymmetric mode only emerges above a band gap at $\approx 3.5\;$GHz. This band structure explains the main features of the observed transmission: Put into line $1$ the wave is not in an eigenmode and will excite both modes, which then propagate with different (Bloch-)wave vectors $K_{1,2}$, so that a beating pattern in space results (similar to the physics of evanescently coupled waveguides or coherent oscillations in time in a double-well). The observed resonances and anti-resonances are a direct result of the beating, as is the flat transmission region in the bandgap of the anti-symmetric mode, where the input is split symmetrically into direct and coupled port by the symmetric eigenmode.
This simple picture is additionally modified by scattering from the in- and out-coupling into the periodic structure, which leads to small wiggles associated to Fabry-Perot-type resonances, particularly pronounced just above the bandgap (see Appendix~\ref{Sec:FP-resonances}). Other important modifications stem from dissipative effects (although weak), from disorder of the `crystal'-strcuture due to fabrication imperfections (see Appendix~\ref{Sec:disorder}) and from any parasitic resonance.


\begin{figure*}[t!]
\includegraphics[clip,trim=7cm 10.0cm 0cm 9cm,width=1\textwidth]{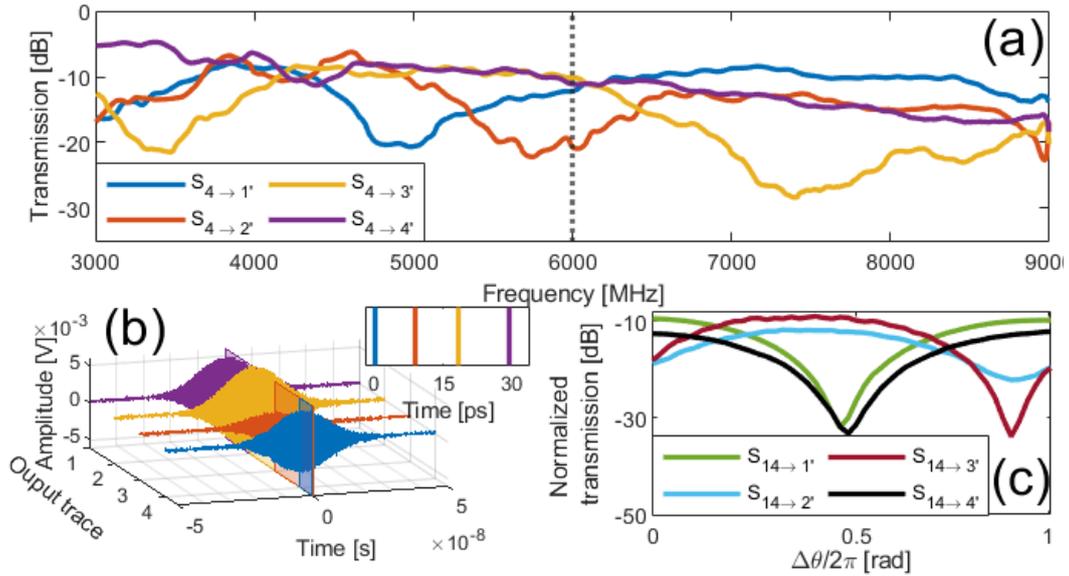}
\caption{\label{fig:Nockit7}Linear transmission measurements of the 7PMT as described in the main text. (a) Transmission spectra for signals introduced in trace no. 4, with rolling averages of 150 MHz to eliminate ringing caused by minor reflections at connectors. The dotted line at 5.99 GHz marks the central frequency of Gaussian wave packets used in pannel (b). (b) Measurement of propagation and arrival times of wave packets with the central frequency $f_4$, introduced in the center waveguide. Perpendicular squares mark the center of the wave packet, corrected for unequal launching traces. Inset: Zoom on the (time, output trace)-plane analogue to the colored planes in the main figure.(c) Measured transmission to ports 1'-4', as we split the CW input signal at $f_{1,4}=5.12$ GHz between port 1 and 4, varying the phase difference.}
\end{figure*}

We also observe interference between signals introduced simultaneously in the two waveguides: In Figure~\ref{fig:Nockit2}(b) we alter the phase difference between the two inputs, and while we measure the transmission through one wave-guide, the signal power in the other ("the neighbor") is scanned over four orders of magnitude (and for all relative phases). 
For the lowest input powers into the neighbor, the direct transmission is drastically reduced due to nonlinearities. This effect is reproduced by our numerical simulation where nonlinearities are present only in the couplers. As the power in the neighbor increases, the signals interfere, and the nonlinearity of the couplers quenches and phase shifts the highest-power signal outputs.

In our next experiment, we increase the network's size to include seven parallel microstrip traces (7PMT) in a circuit similar to the former one, as portrayed in Figure~\ref{fig:Concept}(d), which also visualizes the layers of the fabrication scheme. In this device, we boost the couplers' $Z_l$ further by removing the ground above them (not shown), thus minimizing their $C_l$. The performance is tested by applying CW signals over a bandwidth of $6$ GHz in the center trace (no. 4) and measuring the output, presented in Figure~\ref{fig:Nockit7}(a). Here the dotted vertical line marks the frequency $f_4=5.99$ GHz, chosen as the central frequency of wave packets used for the subsequent measurement. We then replace the CW signal with short Gaussian-shaped wave packets generated by side-band mixing control, again introduced in the center trace. Their arrival is detected at the output terminals of the device. Overall, each wave packet traverses the network in nanoseconds, but when we subtract the electrical delay, we register the arrival at different output traces with a relative delay of $\sim 10$-$30$ ps (see Figure~\ref{fig:Nockit7}(b)), compared to the arrival of the first wave packet at port 4'. In this figure, the smaller amplitude of the detected wave packet at port 3' (shown in red) is consistent with the lower transmission due to interference through that specific trace.
\begin{figure*}[t!]
\includegraphics[clip,trim=0cm 11cm 5cm 8cm,width=1\textwidth]{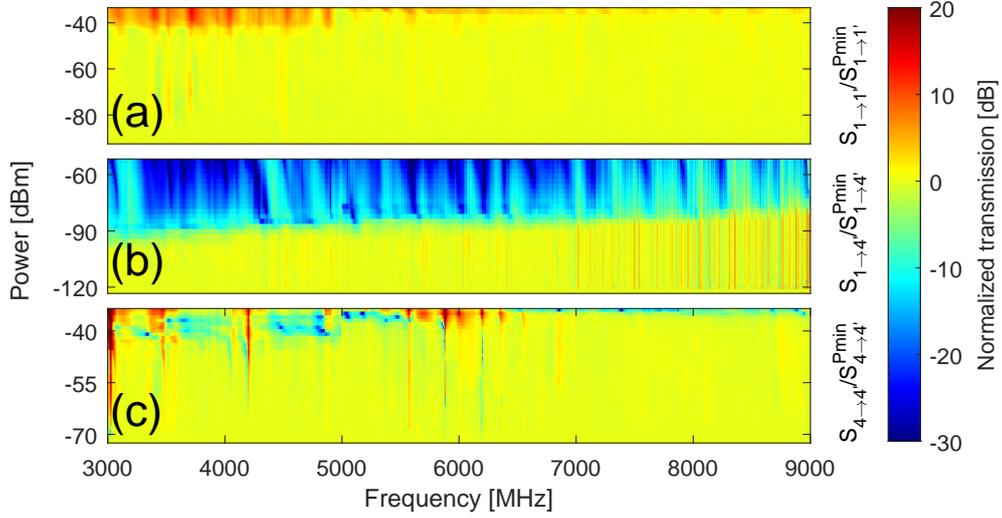}
\caption{Power transmission spectra in the 7PMT measured for varying input powers, (a) through trace 1, (b) from port 1 to port 4', and (c) through trace 4. Labels are the same as in Figure~\ref{fig:Concept}(d)). The color-scale is common for all three subfigures, and indicates the power-dependent transmission normalized to the transmission of the lowest power ($P_{min}$) in the spectrum. 
}
\label{fig:7PMT_nonlin}
\end{figure*}

Returning to CW signals, we proceed at the frequency $f_{1,4}=5.12$ GHz for which the eight transmissions ratios from ports 1 and 4 to ports 1'-4' (according to annotation in Figure~\ref{fig:Concept}(d)) are all relatively high and similar in magnitude. Splitting the input power between the former two, we vary the relative phase and measure the output in Figure~\ref{fig:Nockit7}(c). The nearly symmetrical interference patterns are due to similar transmission coefficients in the network (e.g. $4\rightarrow 1'$ vs. $1\rightarrow4'$). Injection at port 4 has the possibility also to coherently diffuse to traces 5'-7', causing the slight asymmetries in Figure~\ref{fig:Nockit7}(c).

Similar theoretical considerations as for the double-line can be employed for the 7PMT. Besides the band structure and symmetry of the eigenmodes, our simulations (see Appendix~\ref{app:Sim7PMT}) show that propagation through the waveguides resembles quantum walks observed in optical systems \cite{ng2020mapping} with a diffraction pattern, related to the frequency-dependent transmission in Figure~\ref{fig:Nockit7}(a).

We also measure the nonlinearity in the 7PMT emerging from the HKI of WSi by transmitting CW signals of increasing powers through chosen waveguides, starting at signals corresponding to an occupation of $\sim 1$ photons in the device. The frequency-dependent transmissions, plotted in Figures~\ref{fig:7PMT_nonlin}(a)-(c), clearly show that the nonlinearity first emerges in the couplers before it manifests in the waveguides. The direct transmission $S_{1\to 1'}$, is thus hardly affected until the highest excitations are reached and the signal is confined in the trace. Transmitting power from this waveguide to the center of the device relies on couplers between all waveguides in between, resulting in the stronger power dependence of $S_{1\to 4 '}$. The case of transmission through the central waveguide (trace 4) differs from the two above: Despite again considering a coupler-free transmission path, this waveguide is coupled on either side and therefore is more sensitive to the couplers’ behaviour. These effects are further discussed in Appendix~\ref{app:Sim7PMT} (and in its related Figure~\ref{fig:powerdistr}).

\begin{figure*}[t!]
\includegraphics[clip,trim=-3cm 0cm -3cm 0cm,width=1\textwidth]{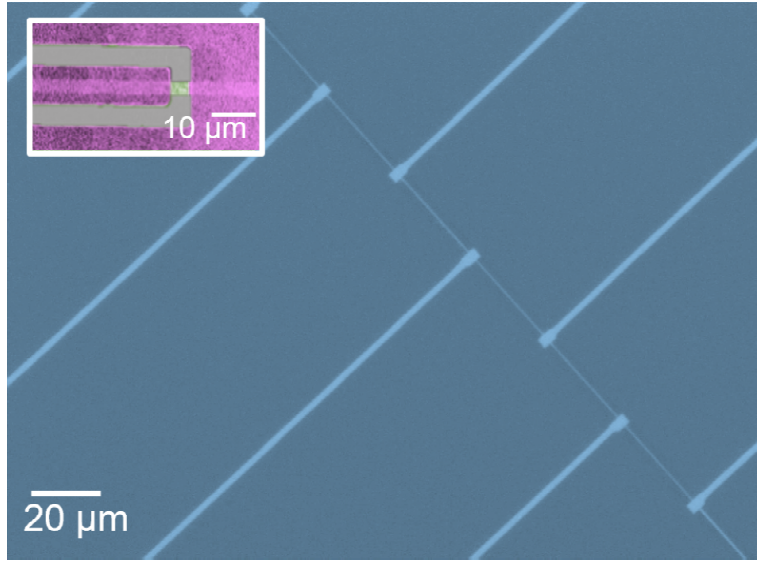}
\caption{Optical microscopy image (false colored), showing parts of wide microstrip resonators coupled with narrow couplers. Inset: SEM photo of coupling to input or readout with colors matching Figure~\ref{fig:Nockit7} (grey substrate and green WSi visible in the gap separating purple Al readout line from purple ground).\label{fig:2DSL_microscope}}
\end{figure*}

\subsection{Resonant Cavity of Standing Waveguides}

\begin{figure*}[t!]
\includegraphics[clip,trim=-0.5cm 10cm 1.5cm 9cm,width=1\textwidth]{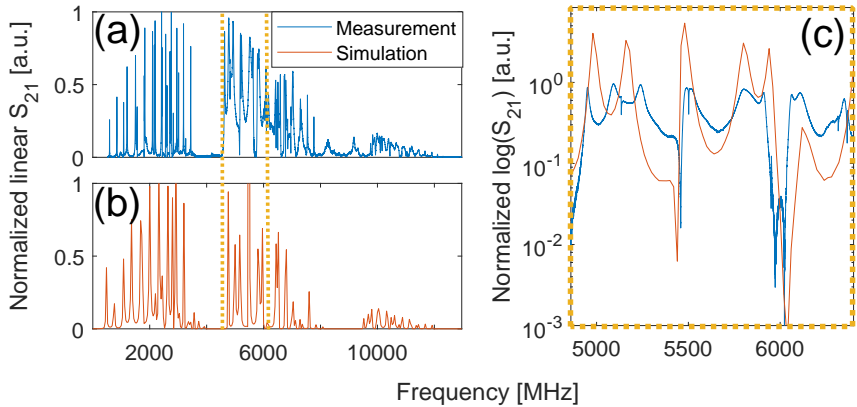}
\caption{Linear transmission vs. frequency normalized by the the strongest response shown by (a) measurement and (b) simulation. For this device there are only two ports, 1 and 2. Here and elsewhere $S_{x\to y}=P_y/P_x$, where $P_i$ is the power at port $i$. (c) Zoom on part of the spectrum joining a and b.\label{fig:Lockit}}
\end{figure*}

The third and final demonstration of the capabilities of superconducting microstrip WSi circuitry switches the focus from traveling to standing waves in a 2D square lattice (2DSL) of 49 microstrips, effectively acting as a multi-mode resonance cavity. Each microstrip resonator is $\sim 400$~\textmugreek m long and is coupled to four neighbors (two in either end, shown in Figure~\ref{fig:2DSL_microscope}). The resonators in the two opposing corners of the 2DSL are capacitively coupled to coplanar transmission lines (inset of Figure~\ref{fig:Lockit}(a)), terminated in large ($0.3$ mm wide) launch-pads, enabling excitation and measurement. Scanning CW the transmission spectrum (Figure~\ref{fig:Lockit}(a)) reveals three distinctive energy bands within the operational bandwidth of our readout-chain, comparable to the linear simulation in figure~\ref{fig:Lockit}(b), which considers both dielectric loss and the transmission profile of attenuators, amplifiers, and circulators applied in the experiment. The simulation, analogous to that in Figure~\ref{fig:Nockit2}(a), also correctly reveals finer features within the energy bands (Figure~\ref{fig:Lockit}(d)), and shows the band structure to be largely determined by the couplers. When these are longer, bands and gaps are dense, as modes populate the couplers. In the opposite limit, reduction of the coupler length breaks down the well-ordered band structure. 
Importantly, our measurements span several orders of magnitude in power, starting from $P=-120$ dBm, which corresponds to an expectation of $0.2$ photons within our device (given by $PL/(hfv_{ph})$, with $h$ Planck's constant, $f=6$ GHz, and $L=3$ mm for the 7PMT traces). Remarkably, the nonlinearity of the couplers confines the transmission in Figure~\ref{fig:7PMT_nonlin}(b) at $-90$ dBm corresponding to only 200 photons. 

The 2DSL's geometry is closely related to that of photonic gratings employed to demonstrate a variety of many-body problems \cite{hartmann2008quantum}, such as quantum entanglement \cite{raimond2001manipulating}, interacting polaritons \cite{hartmann2006strongly}, and phase transitions of Mott-Insulators \cite{jaksch1998cold,larson2008mott}.

\subsection{Estimating the Kerr nonlinearity}
\label{subsec:Kerr}
\begin{figure*}[t!]
\includegraphics[clip,trim=0cm 10cm 0cm 10cm,width=1\textwidth]{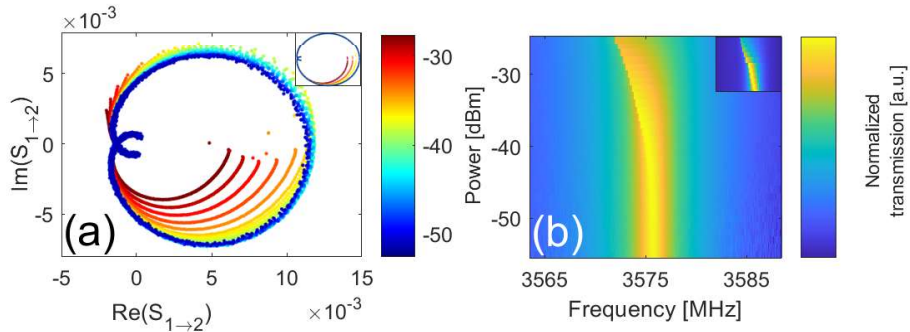}
\caption{(a) Nonlinearity of the 2DSL measured by a polar representation of the transmission $S_{21}$ in one of the peak frequencies from Figure~\ref{fig:Lockit}(b) (inset: simulation) and (b) magnitude of the transmission around the same peak frequency as shown in (a) (inset: simulation).} 
\label{fig:2DSL_nonlin}
\end{figure*}

Nonlinearity is observed in the 2DSL, when we introduce sufficiently strong powers and it affects the resonance frequencies' phase and magnitude (exemplified in Figures~\ref{fig:2DSL_nonlin}(a) and (b) respectively). The observed behavior can be explained by a Duffing-type toy-model of a Fabry-Perot resonator (see theory results in the insets), where the phase accumulation, when crossing the mirrors and the cavity itself, is assumed to become power dependent (see Appendix \ref{AppendixDuffing}).

The observed power dependence is quantified as a self-Kerr nonlinearity \cite{yurke2006performance}, and is approximated as the linear shift in frequency per additional photon, i.e. \begin{equation}
    K_{11}\sim \frac{\Delta \omega}{\Delta N} =\frac{2\pi(f_1-f_0)}{N_1-N_0}
\end{equation}
where $f_{0,1}$ are the resonance frequencies at two different powers, and $N_{0,1}$ the corresponding number of photons in the cavity. The frequency dependence on the photon number is estimated by means of the Q-factor:
\begin{equation}
    N=\frac{2P}{\hbar \omega}\frac{Q}{\omega}
    \label{eq:Qfactor}
\end{equation}

where $P$ is the power. The first fraction in Equation~\eqref{eq:Qfactor} is the rate of photons entering the resonator, and the second fraction is the average survival time of a photon. 

For the resonance shown in Figure~\ref{fig:2DSL_nonlin} at powers of -$40$ dBm and $-55$ dBm, we find $K_{11}\simeq-7.8\times10^{-4}$ Hz. The frequency decreases, when photons are added, so $K_{11}$ is negative, but its magnitude is remarkably larger than the corresponding results found for $w=8$~\textmugreek m \cite{kirsh2021linear}, consistent with $K_{11}\propto1/L_{kin}^2\propto 1/w^2$, cf. Equation~\eqref{eq:Lkin}. Note that the investigated resonance is an extended mode residing in a network of multiple coupled resonators. The mode volume is significantly enhanced and the nonlinearity is therefore somewhat suppressed. When a single resonator is probed, the nonlinearity can be $\sim2$ orders of magnitude larger \cite{kirsh2021linear}.

\subsection{Dielectric losses}
The use of a thin dielectric barrier for the microstrip capacitance leads to losses from two-level-systems (TLSs) in the dielectric material \cite{kirsh2017revealing,olivier2003coupled,tamura2006microwave}. Powers above a certain material-dependent threshold saturate the TLSs, and the resulting transmission spectrum reflects the nonlinearity of the dielectric rather than that of the waveguide. This effect can be roughly estimated quantitatively by the saturation parameter, a function of the TLS Rabi frequency \cite{steck2007quantum}. 

However, the short length (up to 10's of wavelengths) of the itinerant devices ensures minimal losses ($< 10$ \%), when using a low loss-tangent ($<5\times10^{-4}$) barrier material such as amorphous Si. In future designs an alternative dielectric could replace amorphous Si to allow even higher transmissions and thus signals closer to the single-photon-limit.

\section{Outlook}
\label{sec:Outlook}
In this work, we have introduced a platform for on-chip microwave photonic experiments with superconducting circuits. Our three devices, fabricated with established cleanroom procedures, utilize the HKI of WSi in a microstrip geometry. This property, together with the strongly reduced phase velocity, allows us to demonstrate rich phenomena of linear and non-linear optics in on-chip impedance-matched networks of coupled microwave transmission lines.

The three setups presented here constitute first examples of
functionalized devices in this platform. They are chosen to demonstrate
the possible design versatility that can be advanced to future devices
with greater functional complexity.
The first setup realizes the simplest linear optics device, a
beam splitter \cite{gabelli2004hanbury}, by replacing the wave coupling of
similar devices in integrated optics
\cite{o2009photonic,wang2020integrated} by periodic couplers. The crystal like-structure enriches the design variability by the ability to employ band-structure design techniques, for instance, with the aim of creating photonic band gaps or other device principles
from photonic crystal or semiconductor physics. In addition, we demonstrated strongly nonlinear effects in the CW propagation. Secondly, we extended the double-line device towards a more
complex network, which mimics multi-scatterer configurations used for boson-sampling in quantum optics experiment in the visible regime. There, we studied power diffusion between the traces, pulse propagation, and nonlinear effects.

Waveguide lattices of similar type may also be used for (microwave)
photonic simulations, while nonlinearities can be exploited for
wave-mixing and non-reciprocity \cite{kamal2011noiseless}. Finally, in the third setup we realized a network of weakly coupled resonators with multiple pronounced resonances. The Duffing-like nonlinear transmission through one such resonance was
investigated in detail.


Going beyond the simplest linear optics devices, the platform presented here will be able to implement both linear and nonlinear functional units, either without (passive) or with (active) external parameter modulation. For instance, passive linear devices could be built exploiting band structure design to create low or high pass filters, or by designing destructive interference to achieve zero transparency. Our design flexibility in terms of device geometry can be used to build loop resonators, or side-coupled stub resonators, to shape Fano resonances or other desired transmission profiles \cite{limonov2017fano}. Combined with non-linearity, such devices have all the ingredients for nonreciprocal effects and can be used to design diodes. Other possible nonreciprocal units are active devices, e.g. parametrically driven, which can be applied as routers and circulators \cite{kamal2011noiseless}. In our platform, such devices can be realized by nonlinear frequency mixing with the signal in a control port of a multi-port geometry, or by direct modulation of linear devices parameters.

While the working principle of these integrated optics devices rely on classical wave physics, subject to modifications, our platform may also find use in scattershot boson-sampling \cite{latmiral2016towards}, multi-mode few-photon interferometry \cite{goldstein2017decoherence}, for analogue simulation of effects such as Hawking radiation \cite{nation2009analogue}, or as the non-linear medium exploited for reservoir-computing in neural networks \cite{angelatos2020reservoir}.

\section{Acknowledgements}
We acknowledge the support of ISF grants 963.19 and 2323.19 and of the DFG Grant No. AN336/13-1, the IQST, and the Zeiss Foundation.
\appendix
\section*{Appendices}
\setcounter{subsection}{0}
\section{Estimating the Phase Velocity}
An important property of our microstrip networks is the phase velocity, $v_{ph}$ which depends on frequency and geometry of the traces. For the couplers, the geometric inductance, $L_{g,l}$ is completely negligible, and in the linear regime Equation~(\ref{eq:Lkin}) reduces to
\begin{equation}
L_l=\frac{\mu_0 \lambda_L^2}{(t\cdot w)},
\end{equation} where $w$ and $t$ are the width and thickness of the trace in question (written explicitly instead of $A$). But also $C_l \propto w$, so for traces with submicronic ranges, $v_{ph}=(C_lL_L)^{-1/2}$ is independent of $w$. However, in wider traces such as our wave guides, $L_{g,l}$ becomes important, hence raising the total $L_l$ somewhat, and lowering $v_{ph}$ by $\sim 10\%$. Wave-guides and couplers are thus foremost differentiated by their impedance $Z$.

We estimate $\lambda_{WSi} \sim 450$~nm for our sputtered $W_{0.55}Si_{0.45}$ based on other measurements (not shown here), which is in the same order of magnitude, but moderately lower than more tungsten-rich alloys \cite{zhang2016characteristics}.

Our measured $v_{ph}=4\times10^6$ m/s fits its theoretical value found using the formulae and values in this section, and the results also agree with the computed microstrip $v_{ph}$ from \cite{mazin2010thin}.

\section{Technical Aspects of Fabrication}
\label{Sec:TechnicalFab}
In Section \ref{sec:Fabrication} we outlined the fabrication scheme's various steps associated with the three layers of our device. Here we include additional technical details.

After WSi deposition, the applied photo-resist is AZ1505, spun at 4000 RPM. Exposure with a 405 nm laser is followed by development in AZ developer for a minute, and prior to wet-etch, we hard-bake our devices at 120$^\circ$C for 2 minutes. The etching is done with a tungsten etchant at 3 nm/s (verified in separate experiments), and stopped by immersion in water. 

Narrowing the couplers' width to below the wavelength of the laser-writer includes, as mentioned, electronic lithography. A protective mask of PMMA is spun at similar parameters as above, baked at 160$^\circ$C and exposed at 5 A current and 1600 \textmugreek C/cm$^2$ in a pattern of two large rectangles distanced by the desired coupler over each intended coupler strip (which after the former step was $>$~\textmugreek m wide). The \textbf{un}exposed strip between these openings in the mask are is centered above the strip to be narrowed, and after development in an MIBK solution the process is completed by an additional wet-etch session.

The dielectric Si is grown as we melt and evaporate bulk Si grains by e-beam. The relatively slow evaporation rate (compared to, e.g. the growth rate of Al, mentioned below) as given in the main text, results in the amorphous surface with the desired dielectric constant. Patterning of the Si layer is unnecessary; the WSi and the overlying Al must be in galvanic contact only at the launcher pads, and the large areas of these two layers ensures a sufficiently high capacitance, and in turn a negligible impedance $Z_C \ll (i\omega C)^{-1}$ for  the range of frequencies in our measurements.

In developing the fabrication recipe, we tested two methods for patterning of the Al ground: Sputtering followed by wet-etch (Al etchant) and lift-off of an evaporated Al film. The advantage of the former is the high quality and uniformity of a sputtered metal film, but this method included alignment through the highly opaque Al layer, when exposing the spun photoresist, intended to serve as a protective mask during the etching step.

The alternative, lift-off, includes a two-layer mask. Initially, LOR 5B is spun (rates as above) and baked at 200$^\circ$C for 5 min, and subsequently AZ1505 is applied, spun, and baked, and the entire wafer is exposed with parameters as given above. No post-bake is necessary, and we deposit a $\sim$~60~nm film by e-beam evaporation at 0.5 nm/s. 

\section{Fabry-Perot Resonances}
\label{Sec:FP-resonances}
The symmetry of the double-line device with respect to exchanging primary and secondary line is reflected in the (anti)symmetric eigenmodes. If the lines were fed by a symmetric combinations of incoming waves, these would couple to the symmetric eigenmode and result in symmetric outgoing waves. Following this reasoning, we can decouple the double-line into two independent single-channel problems: after introducing (anti)symmetric combinations  Eqs. 10 and 11 result in 
\begin{subequations}\label{a/s_coupling}
\begin{align}
& t^l_+ = \frac{t_p^l + t_s^l}{2} =  \frac{1}{1-z_0^2}   
\Bigl[
a_2 e^{i K_2 l}  \left(1-z_0 e^{i K_2}  \right)  
+b_2 e^{-i K_2 l}  \left(1-z_0 e^{-i K_2}  \right)  \Bigr]  \label{s_coupling}\\
& t^l_- = \frac{t_p^l - t_s^l}{2} =  \frac{1}{1-z_0^2}   
\Bigl[
 a_1 e^{i K_1 l}  \left( 1-z_0 e^{i K_1}  \right)  
+ b_1 e^{-i K_1 l}  \left(1-z_0 e^{-i K_1}  \right)   \Bigr] \label{a_coupling}
\end{align}
 \end{subequations}
and equivalent expressions for  $r^l_{\pm}$, so that we indeed arrive at two decoupled single-channel scattering problems.

\begin{figure}[t!]
\begin{center}
\includegraphics[clip,trim=7.5cm 2cm 8cm 1cm,width=1\textwidth]{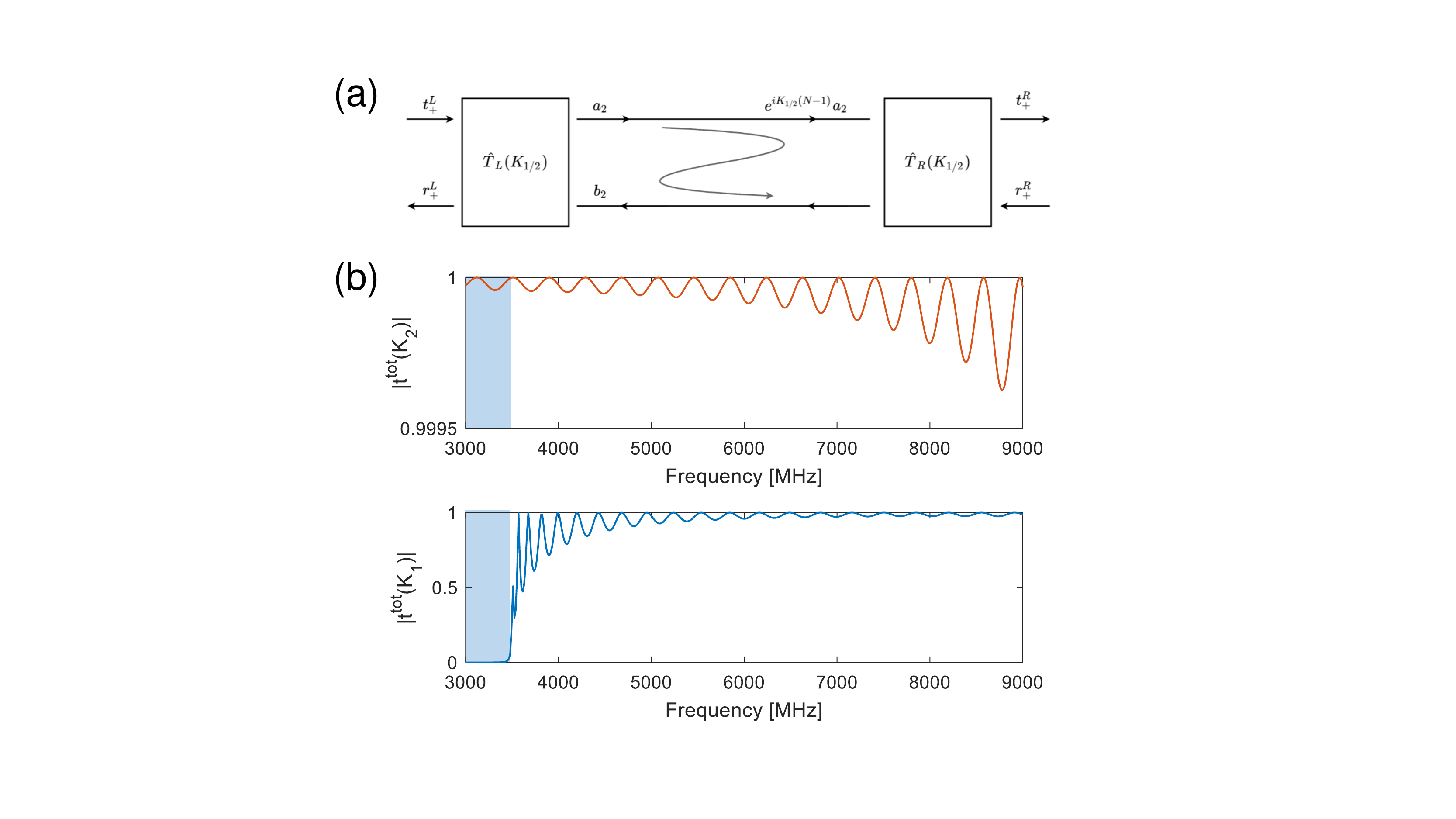}
\caption{(a) Fabry-Perot resonator formed by scattering at in-/out-coupling and propagation in eigenmode of infinite line.
Transmission of symmetric (b) and anti-symmetric eigenmode through the Fabry-Perot structure, cf. Eqs.~\eqref{t_tot_result} and \eqref{eq:FP}.
}
\label{fig:FP}
\end{center}
\end{figure}

To understand results, it is instructive to view the single-channel problem as a Fabry-Perot type scattering problem, where in- and out-coupling constitute a left and right scattering barrier of a resonator, in which propagation is described by the eigenmode, see Figure~\ref{fig:FP}(a). Transfer matrices of the individual barriers, $\hat{T}_{L/R}(K_{1/2})$, are then obtained from Equation~\eqref{a/s_coupling} (and the corresponding equation for $r^l_{\pm}$) and Equation~\eqref{t_tot_result} is recovered from the standard picture of multiple reflections
\begin{equation}\label{eq:FP}
t^\textrm{tot}(K) = \tilde{t}_L e^{i K (N-1)} \left( 1 + \tilde{r}_R e^{i 2 K (N-1)} \tilde{r}'_L + \hdots \right) \tilde{t}_R =
\frac{  \tilde{t}_L e^{i K (N-1)}  \tilde{t}_R } {1 -  \tilde{r}_R e^{i 2 K (N-1)} \tilde{r}'_L }\;,
\end{equation}
where $\tilde{t}_{L/R},\,\tilde{r}_{L/R},\,\tilde{t}'_{L/R},\,\tilde{r}'_{L/R}$ are entries of the scattering matrices corresponding to $\hat{T}_{L/R}(K_{1/2})$.

This picture allows us to explain the features observed in the total transmissions of the symmetric and antisymmetric single-channel problem shown in Figure~\ref{fig:FP}(b).  
In the symmetric case, where the eigenmode wavevector $K_2/L \approx k$ (cf. Figure~\ref{fig:double-line_defs}(b)), in- and out-coupling occurs with minute reflections, so that the total transmission $t^\textrm{tot}(K_2) \approx 1$ with tiny Fabry-Perot oscillations determined by the $e^{i 2 K_2 (N-1)}$ phase factor in the denominator. In the antisymmetric case, below the bandgap  (cf. Figure~\ref{fig:double-line_defs}(b)) total transmission is completely suppressed, while above the bandgap large reflection at the 
in- and out-coupling 'barriers' yield pronounced anti-resonances, which become reduced as $K_1$ grows to approach $kL$. The frequency of oscillations is related to the slope of the $\textrm{Re}\,K_1(\omega)$ curve in Figure~\ref{fig:double-line_defs}(b). 

The total transmission involving the excitation and interference of both eigenmodes is easy to understand below and far above the bandgap:
In the bandgap of the antisymmetric solution, where $\textrm{Im} K_1 >0$, sizeable transmission only occurs through the symmetric eigenmode with $|t^\textrm{tot}(K_2)|\approx 1$ and, hence, $|t^R_{p/s}|\approx 1/2$.
Far above the bandgap, both eigenmodes transmit near perfectly in a wide frequency range and alternately interfere constructively and destructively in primary and secondary line, where the frequency of this interchange is determined by the difference in $K_1-K_2$ stemming from the $e^{i 2 K_{1/2} (N-1)}$ phase factors in the numerators of  Equation~\eqref{eq:FP} resulting in the large-scale structures in the transmission shown in Figure~\ref{fig:Nockit2}(a).
Just above the bandgap, substantial interference can only occur, when the antisymmetric transmission peaks due to a Fabry-Perot resonance, but it will also depend on the respective phases. These resonances are closely spaced (cf. Figure~\ref{fig:FP}(b)), and the result is the rather complex transmission pattern between $3.6$ and $4$ GHz in Figure~\ref{fig:Nockit2}(a).
Similar considerations as for the double-line can be employed for multi-line setups, but there, beyond the band structure and some symmetry considerations on the eigenmode structure, an intuitive understanding becomes considerably harder.

\begin{figure}[t!]
\includegraphics[clip,trim=1cm 10cm 1cm 10cm,width=1\textwidth]{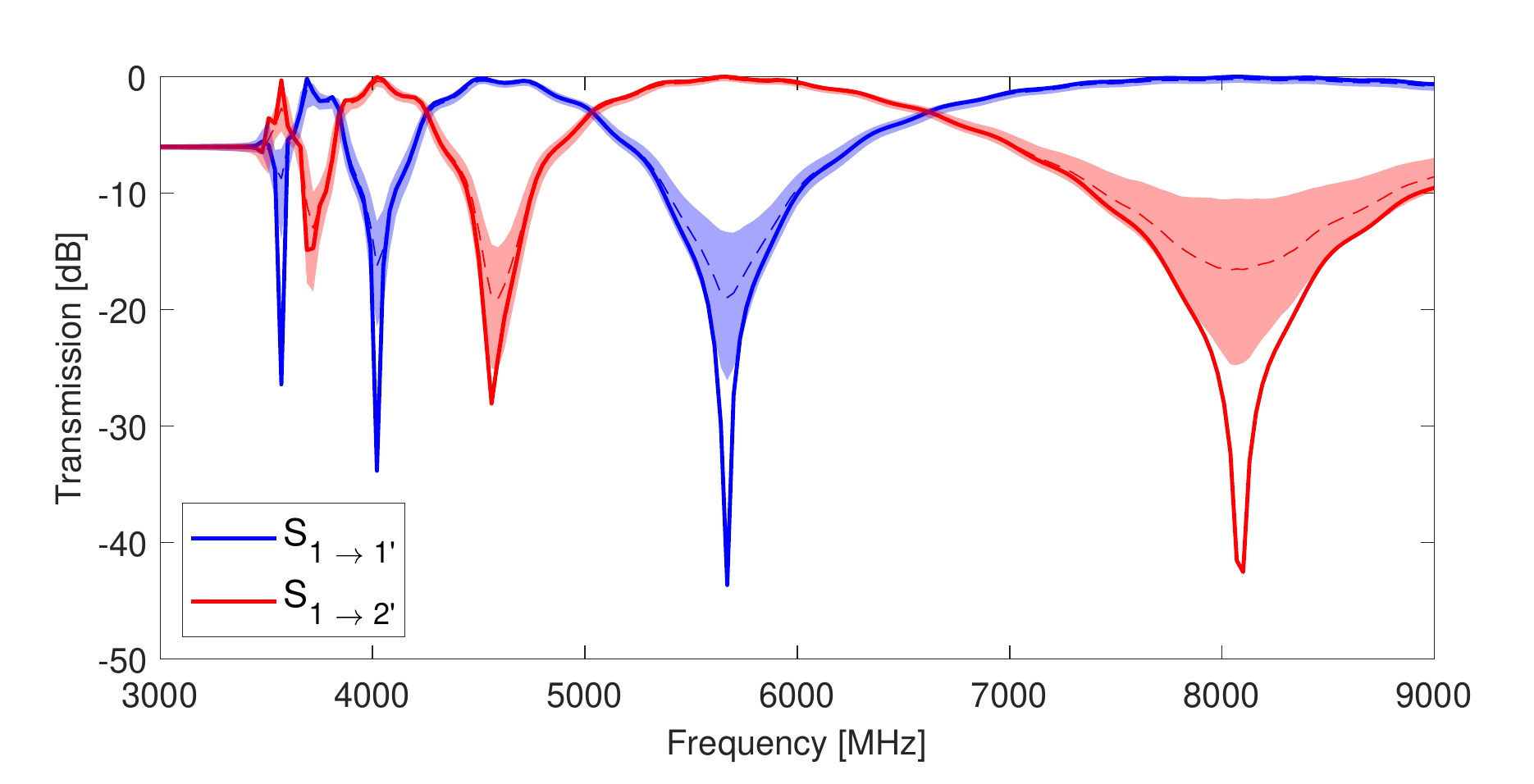}
\caption{\label{fig:disorder}Influence of fabrication imperfection. Direct and coupled transmission through a double-line device without (solid) and with variation of parameters between individual segments.
}
\end{figure}

\section{Disorder of the Periodic Structure}
\label{Sec:disorder}

For all the simulations presented in this work, we assumed devices to have strictly identical parameters for various segments; i.e., for the double-line device all couplers are assumed to have identical length, impedance and capacitance per length, and are equidistantly placed and primary and secondary line are similarly identical. Fabrication imperfections will obviously disturb these symmetries; both the discrete translational symmetry by one unit cell (periodicity) and the p/s-mirror symmetry. As these symmetries were crucial in explaining the experimental measurements, we study the robustness of observed features against disorder in the parameters of the individual segments. For that purpose, we assume independently, normal distributed parameters for inductance, capacitance and length of each individual segment with a relative variance 3\%. Figure \ref{fig:disorder} shows the simulation result for direct and coupled transmission (cf. Figure \ref{fig:Nockit2}(a)) obtained by simulating $N=1001$ such imperfect devices compared to the device without variations (solid lines). The shaded region indicates a $1\sigma$ confidence interval (i.e., for a certain frequency only 16\% of devices fall below (above) the lower (upper) limit) around the median (dashed). Note, that the depth of the destructive interference minima is very sensitive to the symmetry breaking caused by disorder, while other features are relatively robust. 

\section{Simulation of Power Propagation in the 7PMT}
\label{app:Sim7PMT}
\begin{figure*}[t!]
\includegraphics[clip,trim=0cm 2.5cm 0cm 3cm,width=1\textwidth]{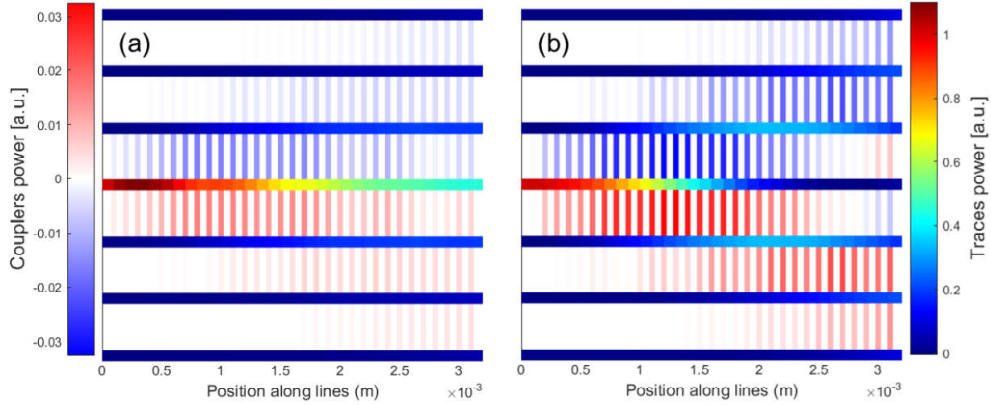}
\caption{\label{fig:powerdistr}Simulations of power distribution in the 7PMT, when introducing signal at $8.5 GHz$ in the central trace at (a) $-40$ dBm and (b) $-90$ dBm. Colorbars include direction (positive is defined downwards and rightwards). Left(right) colorbar relates to couplers(traces) in both subfigures. Both figures are normalized according to the input power (i.e. leftmost cell in the central trace equals unity).}
\end{figure*}
In the main text we discussed the simulated propagation of power throughout the network, both directly through various traces in the linear regime, and in the nonlinear regime. In Figure \ref{fig:powerdistr} we expand the simulation to show the propagation, also in the narrow couplers, too numerous to be shown in the main letter.

In particular, Figure~\ref{fig:powerdistr}(a) displays the delay of transmission from the central trace, where the power is injected, to its neighbors. This is due to the high nonlinearity experienced by most of the couplers linking this trace to the rest. In \ref{fig:powerdistr}(b), where the introduced power is five orders of magnitude lower, fewer couplers are affected and the power transmits to adjacent traces earlier.

\section{Simulation of the Nonlinear Behavior of the Double-Line}

In Section~\ref{sec:ResultsDiscussion} of the main text we discussed the interference between two inputs in the double-line device and showed the impact of nonlinear effects at higher power in Figure~\ref{fig:Nockit2}(b). Here, we will briefly describe, how to use a nonlinear single-frequency simulation to model such effects.

\begin{figure*}[t!]
\includegraphics[clip,trim=0cm 0cm 0cm 0cm,width=1\textwidth]{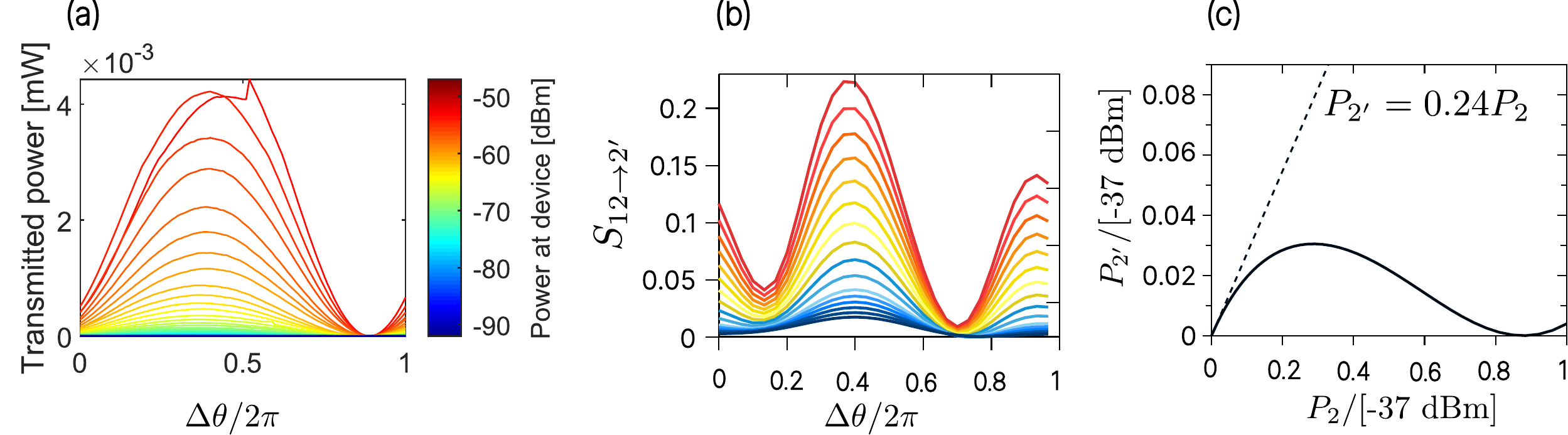}
\caption{\label{fig:nonlinear_interference}
Measurement and simulations illustrating the non-linear behavior of the first device (with two parallel microstrips) in the interference measurement presented in Figure~\ref{fig:Nockit2}(b). Signals are sent down both traces 1 and 2, with constant phase and variable power in trace 1, while in trace 2 the power is kept constant ($P_2\simeq -37$ dBm) and the phase $\theta$ is varied. (a) Measurement curves showing the power output at trace 2' (opposite of trace 2) for different values of the power in trace 1. Non-linear effects manifest as a shift of the phase dependence and jumps that signal multi-stable states. (b) The simulation captures the shift of the phase dependence with increasing power in trace 1. (c) The transmitted power $P_{2'}$ has a linear regime as a function of the input $P_2$ (with $P_1$ set to zero) only up to a power $P_2\simeq -50$ dBm. The measurement $P_2\simeq -37$ dBm corresponds to the non-linear regime, near the zero of $P_{2'}$, where the transmission is drastically reduced, as can be seen in (a) (the transmission is negligible for $P_1\simeq 0$).
}
\end{figure*}

To model the experimental results of Figure~\ref{fig:Nockit2}(b), we consider a nonlinear inductance as in Equation~\eqref{eq:Lkin} for the couplers only, where nonlinear effects are more pronounced.
In Section~\ref{sec:ResultsDiscussion} of the main text we explained how nonlinearities can be included for signal propagation along certain segments and how this modifies the generic matrix equation \eqref{eq:matrixeq} for a network of arbitrary geometry. The resulting nonlinear problem is solved in Matlab using a Broyden method, within which a standard ODE45-solver is used to find the amplitude relations along each segment. The results of this simulation are shown in Figure~\ref{fig:nonlinear_interference}.

In the experiment we observed at the lowest probed power a transmission with a purely sinusoidal phase dependence. Overall the transmission is drastically reduced as compared to the linear regime of Figure~\ref{fig:Nockit2}(a). For stronger power non-linear effects manifest as modified phase dependence with a pronounced minimum at fixed (power-independent) phase difference, and finally as jumps that signal multi-stable states. While our simulation was not designed to capture multi-stability and possible hysteretic behaviour, it can reproduce some of the observed features such as a modified phase dependence and pronounced minima at fixed phase. 

The overall low transmission at the lowest probed power is explained by  Figure~\ref{fig:nonlinear_interference}(c).  
It shows the transmitted power $P_{2'}$ as a function of the input $P_2$ (with $P_1$ set to zero) up to the power of $P_2\simeq -37$ dBm used as the constant reference power in Figure~\ref{fig:nonlinear_interference} (a,b) and the experiment. 
The transmitted power $P_{2'}$ has a linear dependence on the input only up to a power $P_2\simeq -50$ dBm. The measurement $P_2\simeq -37$ dBm thus corresponds to a strongly non-linear regime, namely an input power near the zero of $P_{2'}$. For Figure~\ref{fig:nonlinear_interference} (a,b) this means that for the lowest curves (where the power input into $P_1$ is non-zero but small), the transmission is hence drastically reduced and strongly modulated by the phase of the weak $P_1$.

\section{Simulation of a non-linear resonance in the 2DSL}
\label{AppendixDuffing}

The resonance studied in Figure~\ref{fig:2DSL_nonlin} is modeled as a resonance of a toy-model Fabry-Perot cavity with Duffing non-linearity, acting as proxy for the much more complex resonant structure realized in the experiment.
The output $a_\text{out}$ amplitude of a Fabry-Perot cavity is related to the input $a_\text{in}$ at one of the cavity mirrors (port $1$) by the total transmission amplitude, $a_\text{out}=S_{21} a_\text{in}$, which depends on the 
transmission amplitudes $t_1$ and $t_2$ and reflection amplitudes  $r'_1$ and $r_2$ of the two scatterers (mirrors) defining the cavity:
\begin{align}
S_{21} = \frac{|t_1||t_2| e^{i\phi}}{1-|r'_1| |r_2| e^{2 i K_\text{eff} L}}\,,
\end{align}
where we have assumed a symmetric cavity,  $|t_1|=|t_2|=t$ and $|r'_1|=|r_2|=\sqrt{1-t^2}$.
Important to the model are the phases $2 K_\text{eff}L$ and $\phi$. The first models the phase accumulated during one round-trip through the cavity, while in the second includes contributions from passing through the mirrors. Both phases are assumed linear in the input frequency $\omega$ in the linear regime and account for a Duffing-type non-linearity by including a power dependence (where $|a_\text{out}|^2$ stands in for the intra-cavity intensity),
\begin{align}
\phi & =  \frac{(\omega-\omega_\text{res})}{\alpha \gamma}\left(1-\beta \frac{\gamma}{\omega_\text{res}} \frac{|a_\text{out}|^2}{P_c}\right),\\ 
K_\text{eff} L & =  2\pi \frac{\omega}{\omega_\text{res}} \left(1+\frac{\gamma}{\omega_\text{res}} \frac{|a_\text{out}|^2}{P_c}\right).
\end{align}
Here we have introduced parameters that characterize the resonance: frequency $\omega_\text{res}$, linewidth $\gamma=t^2\omega_\text{res}$, and critical output power $P_c$ above which the Duffing curve becomes multiple valued. These can be directly extracted from the measurement. The two numerical parameters $\alpha$ and $\beta$ are the only fitting parameters, that are found to be of order unity. ($\alpha=2$ and $\beta=1.6$). With this model, we can reproduce all features of the experimental results for the transmission amplitude $S_{21}$, as shown by the simulation results in the insets of Figure~\ref{fig:2DSL_nonlin}.

\begin{figure*}[t!]
\includegraphics[clip,trim=-3cm 0cm -3cm 0cm,width=1\textwidth]{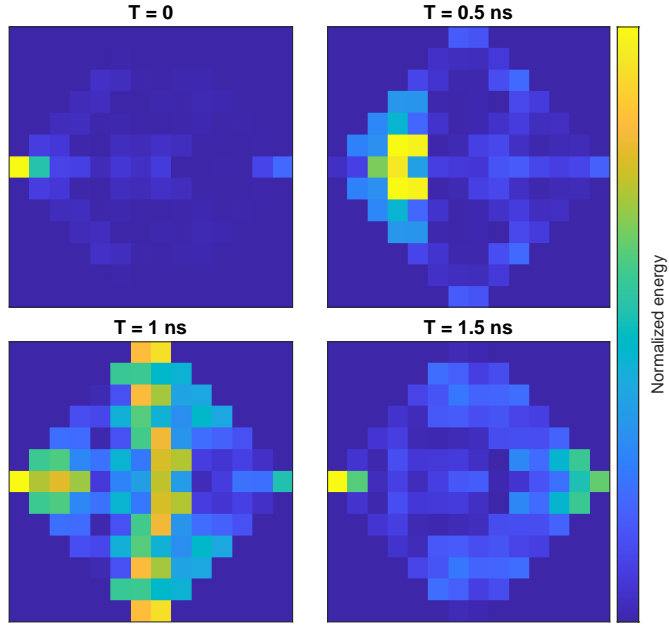}
\caption{\label{fig:ModeExpansion}Simulated power propagation of the initial state mainly occupying the left-most resonator. Note that the energy scale in is normalized in each image according to the highest value (which slightly decreases due to loss effects).}
\end{figure*}

\section{Simulation of the Energy Diffusion}
Our simulations elaborate on future experiments possible with a lattice of coupled microwave resonators as the one we produced, though not experimentally possible with the given instantaneous bandwidth of the IQ mixers in our lab. By choosing the 15 highest peaks in the second energy band shown in the main text's Figure \ref{fig:Lockit}(b) (between 4.8 and 8 GHz) and their corresponding wave functions $\vv{\psi}_m$ (presumed to be eigenmodes of the system), we simulate the evolution of
\begin{equation}
  \vv{\chi}(0) =\sum_{m} \alpha_m\vv{ \psi}_m
\end{equation}
where $\vv{\chi}(0)$ denotes the state, where only one of the corner-resonators in the lattice is excited. Ideally, if $\vv{\psi}_m$ was a complete set of orthonormal eigenmodes, we would expect $\vv{\psi_n}\cdot\vv{\psi_n}^*=\delta_{n,m}$. This would imply that
\begin{equation}
    \alpha_m = \vv{\psi}_m^*\cdot \vv{\chi}(0)
    \label{eq:orthogonal}
\end{equation}
In reality, the main text's Figure~\ref{fig:Lockit} shows only modes with considerable energy in the two resonators used for input and output respectively. Thus $\vv{\psi}_m$ does not obey Equation~\eqref{eq:orthogonal}, but approximates it as the unnormalized wavefunctions holds  $|\vv{\psi}_n|^2 \gg \vv{\psi}_n\cdot \vv{\psi}_{m\neq n}^* $. The limitation of this approximation is visible in Figure~\ref{fig:ModeExpansion}(a), where our attempt to excite only the first resonator also leads to weak excitations in other resonators. 
The evolution of the states over time and the propagation of the energy throughout the lattice is given by
$
    \vv{\chi}(t)=\sum_{m} \alpha_m\vv{ \psi}_m e^{-2i\pi f_m}
$
and is depicted in Figure \ref{fig:ModeExpansion}.

\newpage
\bibliographystyle{iopart-num}
\bibliography{RefsNPJ}

\end{document}